\documentclass[aps,prl,epsfigure,twocolumn,superscriptaddress]{revtex4-2}
\usepackage[colorlinks=true,linkcolor=blue,urlcolor=blue,citecolor=blue,pdfusetitle]{hyperref}
\usepackage[utf8]{inputenc}
\usepackage[english]{babel}
\usepackage{amsmath}
\usepackage[caption = false]{subfig}
\usepackage{graphicx,epstopdf}
\usepackage{threeparttable} 
\usepackage{blindtext}
\usepackage[table,xcdraw]{xcolor}
\usepackage{lipsum}
\usepackage{amsfonts}
\usepackage{bbm}
\usepackage{ulem}
\usepackage{amssymb}
\usepackage{enumerate}
\usepackage{color}
\usepackage{latexsym}
\usepackage{physics}
\usepackage{times,txfonts}

\newcommand{\Acal}{\mathcal{A}}

\newcommand{\1}{\mathbbm{1}}

\usepackage[bb=boondox]{mathalfa}

\newcommand{\SubFig}[2]{\ref{#1}{\color{blue}#2}}

\definecolor{bluePoli}{cmyk}{0.4,0.1,0,0.4}
\definecolor{blueGreen}{RGB}{44, 120, 247}
\definecolor{brickred}{rgb}{0.8, 0.25, 0.33}
\definecolor{darkred}{RGB}{204, 0, 0}
\definecolor{darkgreen}{RGB}{0, 102, 50}
\definecolor{darkblue}{RGB}{0, 76, 153}
\definecolor{mygold}{RGB}{255, 128, 32}
\definecolor{mypurple}{RGB}{178, 105, 252}
\definecolor{myorange}{RGB}{204, 102, 0}

\newcommand{\CSIC}{\href{https://ror.org/009wseg80}{Instituto de Física Fundamental}, \href{https://ror.org/02gfc7t72}{Consejo Superior de Investigaciones Científicas}, Calle Serrano 113b, Madrid 28006, Spain}
\newcommand{\QuARC}{Quantum Advanced Research Center (QuARC), \href{https://ror.org/02gfc7t72}{Consejo Superior de Investigaciones Científicas}, Calle Serrano 113b, 28006 Madrid, Spain}

\newcommand{\UFF}{Instituto de F\'{i}sica, \href{https://ror.org/02rjhbb08}{Universidade Federal Fluminense}, Av. Gal. Milton Tavares de Souza s/n, Gragoat\'{a}, 24210-346 Niter\'{o}i, Rio de Janeiro, Brazil}

\newcommand{\UFFSM}{Instituto de F\'{i}sica, \href{https://ror.org/02rjhbb08}{Universidade Federal Fluminense}, Av. Gal. Milton \\Tavares de Souza s/n, Gragoat\'{a}, 24210-346 Niter\'{o}i, Rio de Janeiro, Brazil}

\newcommand{\Title}{Many-Body Localization Induced by Correlated Disorder in Interacting Superconducting Qubits}

\usepackage{etoolbox}

\usepackage{orcidlink}
\begin{document}
	
\title{\Title}

\author{Thiago R. Gir\~ao Souza~\orcidlink{0009-0009-7273-3063}}
\email{thiagorgs@id.uff.br}
\affiliation{\UFF}

\author{Andreia Saguia~\orcidlink{0000-0003-0403-4358}}
\email{asaguia@id.uff.br}
\affiliation{\UFF}

\author{Alan C. Santos~\orcidlink{0000-0002-6989-7958}}
\email{alan.santos@csic.es}
\affiliation{\QuARC}
\affiliation{\CSIC}

\author{Marcelo S. Sarandy~\orcidlink{0000-0003-0910-4407}}
\email{msarandy@id.uff.br}
\affiliation{\UFF}

\begin{abstract}
	The failure of quantum thermalization due to Many-Body Localization (MBL) has evolved from a theoretical concept in spin chains to an experimental reality in synthetic quantum platforms, most notably superconducting circuits based on transmon qubits. Despite its significance, the MBL transition has been studied primarily under purely random disorder and local couplings, leaving more complex and realistic configurations largely unexplored. Here, we investigate the quantum dynamics of transmon networks subject to the unavoidable competing effects of correlated disorder and network-mediated non-local interactions. First, we show how to engineer the physical parameters of the quantum hardware to systematically control the correlated disorder patterns emerging in the system. Then, we demonstrate that the MBL phase transition is robust against such correlations, which is essential for tuning localization properties in realistic transmon devices. This robustness is established through the analysis of the block entanglement entropy variance across disorder realizations, which precisely locates the MBL critical point. Independently, we introduce a local memory parameter, whose dynamics at long evolution times reveals memory retention in the localized phase and yields a critical point consistent with the entropy analysis. These results provide a framework for understanding localization in complex quantum network topologies, with potential implications for multi-qubit processor design.
	
\end{abstract}

\maketitle

\paragraph{Introduction.} 
The phenomenon of thermalization in isolated quantum systems is typically explained by the Eigenstate Thermalization Hypothesis (ETH)~\cite{Deutsch:91,Srednicki:94,Rigol:08}, which implies that local subsystems eventually lose memory of their initial conditions, effectively acting as thermal baths for themselves. The thermalization process scrambles local information across the system's degrees of freedom as a direct consequence of entanglement spreading throughout the quantum dynamics. This mechanism mimics the propagation of information in a decoherent environment, making the initial state recoverable only by measuring global observables~\cite{Nandkishore:15,Abanin:19,Sierant:25,Pawlik:26}. However, localization can prevent thermalization. Indeed, Anderson~\cite{Anderson:58} demonstrated that a random potential halts the transport of non-interacting particles in real space. For interacting systems, this phenomenon persists in the form of many-body localization (MBL), providing a generic route for thermalization failure under sufficiently strong disorder~\cite{Gornyi:05,Basko:06}. 

Although early theoretical work focused on spin chains, specifically disordered Heisenberg~\cite{Pal:10} and Ising models~\cite{Kjall:14}, experimental realizations have since expanded to diverse synthetic platforms, including ultracold atoms in optical lattices~\cite{Schreiber:15} and trapped ions~\cite{Smith:16}. Parallel to these platforms, superconducting circuits, particularly transmon qubit architectures~\cite{Xu:18,Berke:22}, have emerged as a prominent testbed for investigating MBL in highly controllable environments. From a many-body perspective, these architectures function as lattices of coupled nonlinear quantum resonators, where disorder is typically introduced via intentional frequency detuning between neighboring qubits. Strikingly, this engineering choice also underpins the stability of practical quantum processors, which rely on maintaining a localized phase to suppress the onset of quantum chaos and prevent the subsequent loss of quantum information.
Pushing this field to its frontier, here we will consider localization in more complex configurations, featuring correlated disorder and network-mediated non-local interactions. While purely random disorder serves as the standard theoretical starting point~\cite{Bahovadinov:22}, we will show that correlated disorder patterns, which are typically unavoidable in transmon qubit devices,  preserve the localized phase structure, with the MBL phenomenon exhibiting considerable robustness against inter-qubit correlations. This robustness may be particularly relevant for optimizing the performance of quantum hardware in the presence of localization effects.

We will show how to engineer the physical parameters of the quantum hardware to control the correlated disorder patterns emerging in the system. This is achieved by tuning the shunting capacitances, Josephson energies, local external fluxes, and qubit-qubit coupling capacitances, which together establish a mapping between physical hardware elements and the effective spin couplings and fields that appear in standard MBL characterizations. Building on this mapping, we will demonstrate that the critical MBL behavior is maintained by analyzing the block entropy of the energy eigenstates. More specifically, we will show that the block entropy variance over disorder realizations for energy eigenstates in the middle of the spectrum can characterize the localization critical point. This will be observed by the 2nd-order R\'enyi entropy, which can considerably simplify the standard analysis through von Neumann entropy. 
In particular, the experimental evaluation of the R\'enyi entropy can be done through
randomized measurements, which is useful to avoid full state tomography~\cite{Hu:25a}. 
Independently, we will introduce a local memory parameter based on the site occupation number, whose
dynamics at long evolution times reveals memory retention in the localized phase and yields a critical point consistent with the entropy analysis. This is achieved through the derivative of the memory parameter with respect to the strength of disorder, showing that the connectivity of transmon networks, which inherently leads to correlated disorder, keeps the MBL criticality of its disorder uncorrelated counterpart.

\paragraph{Correlated disorder in Transmon qubit arrays.}

\begin{figure}[t!]
	\centering
	\includegraphics[width=\columnwidth]{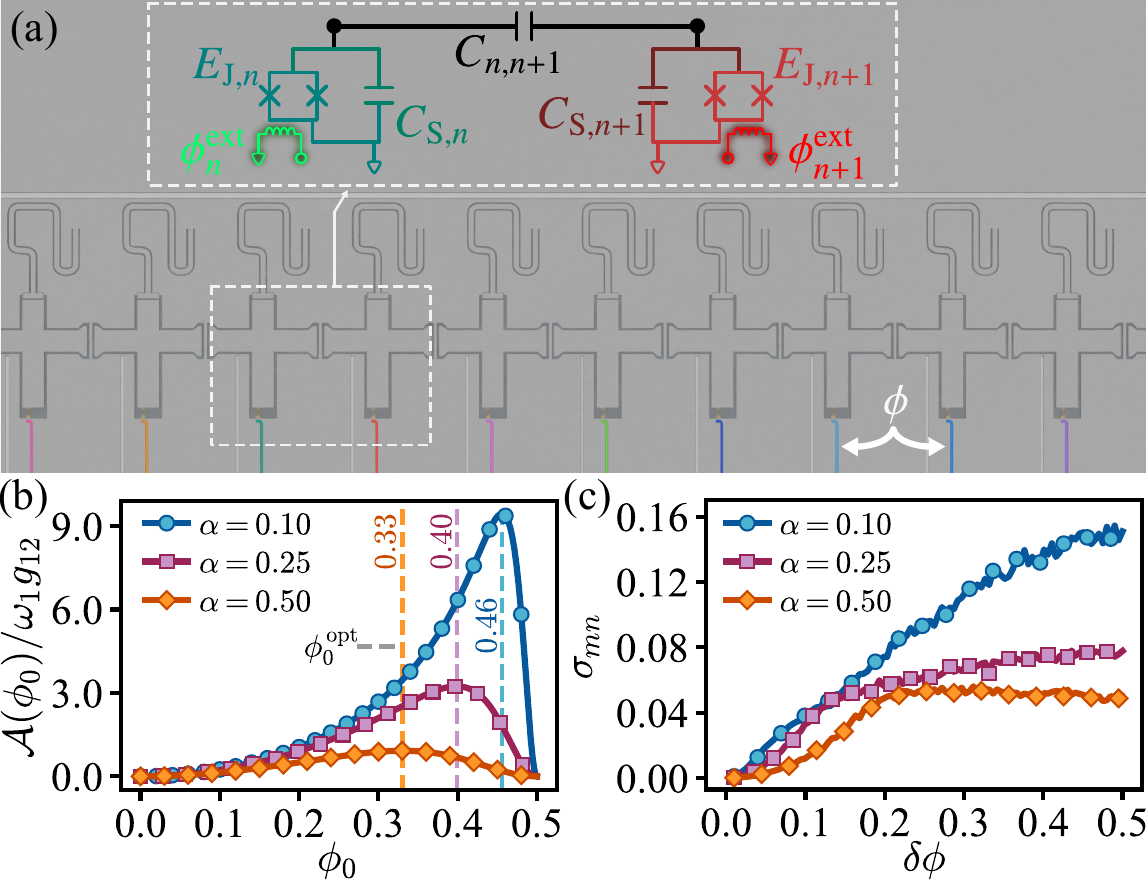}
	\caption{(a) Proposed experimental setup to observe MBL transitions in superconducting circuits through correlated disorder in a linear chain of transmon qubits. Inset we show the circuit diagram for two neighboring qubits, $n$ and $n+1$, with their respective Shunting capacitances ($C_{\mathrm{S},n}$), Josephson energies ($E_{\mathrm{J},n}$), and local external flux ($\phi_{n}$), in addition to the qubit-qubit coupling capacitance ($C_{n,n+1}$). Different colors used for flux lines denote random values of $\phi_{n}$. 
		(b) Coefficient $\Acal_{m}(\phi_0)$, normalized by $\omega_{1}g_{12}:=\omega_{1}(\phi_{0})g_{12}(\phi_{0})$, as function of the reference external flux $\phi_0$ for different values of the junction asymmetry parameter $\alpha$ (shown in (c)). Vertical dashed lines denote the corresponding optimal reference flux $\phi_{0}^{\mathrm{opt}}$ for each value of $\alpha$.
		(c) The covariance is shown as a dimensionless quantity, normalized by $\omega_{1}(\phi_{0}^{\mathrm{opt}})g_{12}(\phi_{0}^{\mathrm{opt}})$. In (b) and (c) we consider a system of $7$ qubits, with experimentally feasible parameters $C_{S,n} = 100$~fF, $C_{n,n+1} = 0.2C_{S,n}$ and $E_{\mathrm{J},n} = 2 \times 10^{-23}$~J~\cite{Ding:23,Wu:24}. The covariance in (c) is evaluated over $3,000$ realizations.}
	\label{fig:scheme-corre}
\end{figure}

The existence of local and network-mediated non-local capacitive couplings in a system of transmon qubits, and their dependence on the Josephson energy and local magnetic flux applied to each qubit enables the emergence of correlated disorder. To demonstrate this result, we consider a device as depicted in \SubFig{fig:scheme-corre}{a}. A system of $N$ transmon qubits is characterized by their Josephson energies $E_{\mathrm{J},n}$, shunting capacitances $C_{\mathrm{S},n}$, and an external local magnetic flux $\phi^{\mathrm{ext}}_{n}$. The transmons are coupled to each other through local nearest-neighbor capacitive coupling, with capacitances $C_{n,n+1}$, created by close proximity of the capacitor pads of the qubits $n$ and $n+1$. In this work we consider a regime of strong capacitive coupling, with $C_{n,n+1} = C_{\mathrm{C}} =0.2C_{\mathrm{S},n}$, unlike conventional transmon devices which usually operate at regimes $C_{n,n+1} \ll C_{\mathrm{S},n}$. Entering this regime is of pivotal relevance to our work because of the emergence of non-negligible non-local interacting terms in the system~\cite{Yanay:22,Rosario:23,Villanova:26}. 

For highly anharmonic transmon artificial atoms, the two-level system approximation can be efficiently implemented for each transmon, with the system governed by the Hamiltonian
\begin{equation}
	\hat{H} = \sum_{n=1}^{N} \hbar \omega_{n} \hat{\sigma}_{n}^{+}\hat{\sigma}_{n}^{-}+ \sum_{n,m < n}^{N} \hbar g_{nm} \left( \hat{\sigma}_{n}^{-}\hat{\sigma}_{m}^{+} + \hat{\sigma}_{m}^{-}\hat{\sigma}_{n}^{+}\right) , \label{Eq:H-XX}
\end{equation}
where $\hbar \omega_{m}(\phi_{m}) = \sqrt{8E_{C,m}E_{J,m}(\phi_{m})} - E_{C,m}$ is the self-qubit energy, $\hbar g_{nm}(\phi_{n},\phi_{m}) = 4E_{C,nm}/\sqrt{4\xi_{n}\xi_{m}}$ is the transmon-transmon coupling, with $\xi_{m} = \sqrt{2 E_{C,m}/ E_{J,m}(\phi_{m})}$, and $E_{C,nm} = e^2 (\textbf{C}^{-1})_{nm} / 2$ is the energy capacitance between the transmons $n$ and $m$ obtained from the inverse of the capacitance matrix $\textbf{C}$ of the circuit device~\cite{Yanay:22,Rosario:23,Villanova:26}. The raising and lowering spin-1/2 operators for the $n$-th transmon in the qubit subspace $\{\ket{g}_{n},\ket{e}_{n}\}$ are defined as $\hat{\sigma}_{n}^{+}= \ket{e}_{n} \bra{g}_{n}$ and $\hat{\sigma}_{n}^{-}= \ket{g}_{n}\bra{e}_{n}$, respectively. The Hamiltonian conserves the total excitation number,	$\hat N_{\mathrm{exc}} = \sum_{i=1}^{L}\hat n_i$, with $\hat n_i=\hat{\sigma}_{n}^{+}\hat{\sigma}_{n}^{-} = \ket{e}_{n}\bra{e}_{n}$, and therefore decomposes into fixed-filling sectors.

Local qubit disorder can be introduced in the system through external field, or by controlling the fabrication process; in both cases the disorder is related to fluctuations in the Josephson energy $E_{J,m}(\phi_{m})$. As a consequence, even small variations in the local flux of the qubit $m$ promote correlated disorder between $g_{nm}$ and $\omega_{m}$. In fact, let us consider small fluctuations of the external flux around their mean values, namely $\phi_m = \phi_0+\delta\phi_m$ and $\phi_n
=\phi_0+\delta\phi_n$, where $\delta\phi_m$ and $\delta\phi_n$ are random fluctuations describing the disorder from the reference value $\phi_0$. By assuming that the fluctuations $\delta\phi_m$ and $\delta\phi_n$ are random and uncorrelated---i.e. $\langle \delta\phi_m \delta\phi_n \rangle  = 0$, with denoting the $\langle \bullet \rangle $ statistical average over many realizations of disorder---the covariance between the energy of the $m$ qubit and its interaction with a qubit $n$ is~\cite{SM}
\begin{align}
	\sigma_{nm} \approx \left.\frac{\partial \omega_m}{\partial{\phi_m}}  \frac{\partial g_{nm}}{\partial{\phi_m}}\right\vert_{\phi_m=\phi_0}
	\left\langle (\delta\phi_m)^2 \right\rangle 
	=
	\Acal_{nm}(\phi_0)\left\langle (\delta\phi_m)^2 \right\rangle  .
\end{align}

The function $\Acal_{nm}(\phi_0)$ serves as an amplifying coefficient of the correlated disorder in the system. Under suitable choices of the reference phase $\phi_0$, the coefficient $\Acal_{m}(\phi_0)$  reaches values, and non-negligible regimes of correlated disorder can be engineered even for small fluctuations $\delta \phi_{m}$.

For instance, in Figs.~\SubFig{fig:scheme-corre}{b,c} we assume the transmons are constituted by asymmetric Josephson loops, where the Josephson energy is $E_{J}(\phi) = E_{J,0} [\cos^{2} (\pi \phi) + \alpha^{2}\sin^{2} (\pi \phi)]^{1/2}$~\cite{Koch:07}, where $\alpha \leq 1.0$ is the junction asymmetry parameter for tunable transmons, which sets the tunability range of the qubit, and $E_{J,0}$ is the Josephson energy at zero external flux $\phi = 0$. In this scenario, it is possible to design devices with a high degree of correlation by choosing the asymmetry of the junctions $\alpha$ and the reference flux $\phi_{0}$.


\paragraph{Entanglement signatures of the MBL transition.} 

The MBL transition is intrinsically dynamical and quantum-coherent in nature, an aspect naturally captured by the entanglement of many-body energy eigenstates. This critical behavior is manifested in the half-system block entanglement~\cite{Pal:10,Kjall:14}. At low disorder, the system is ergodic and block entanglement obeys an approximate volume-law (extensive) scaling. As disorder increases, the system is driven into the localized phase, where block entanglement is strongly suppressed. We probe block entanglement via 2nd-order R\'enyi entropy of the half-system reduced density operator; as we show below, the same critical behavior is corroborated by the von Neumann entropy --- see~\cite{SM} for further details. 

Given an energy eigenstate $|\psi\rangle$ and a bipartition of the transmon array into two contiguous halves $A$ and $B$, the entanglement between $A$ and $B$ is quantified by the second-order R\'enyi entropy $S_{2}$ of either block,
\begin{equation}
	S_{2} = -\log_2 \text{Tr}\left( \hat{\rho}_A^{2}\right) ,
	\label{vonNeumann}
\end{equation}
where $\hat{\rho}_A = \text{Tr}_B(\hat{\rho})$ is the reduced density matrices of block $A$, obtained from the state of the complete system $\hat{\rho} = |\psi\rangle\langle\psi|$. Our choice is mainly motivated by the advantage of measuring R\'enyi entropy in experimental setups, over the von Neumann entropy. In fact, $S_{2}$ can be efficiently estimated through randomized measurements~\cite{Enk:12,Elben:18}, and its experimental feasibility has been witnessed in trapped ion~\cite{Tiff:19} and superconducting qubits platforms~\cite{Hu:25a}. Here, we will show that $S_{2}$ can be used as an efficient measure of the MBL phase transition.

\begin{figure}[t]
	\centering
	\includegraphics[width=\columnwidth]{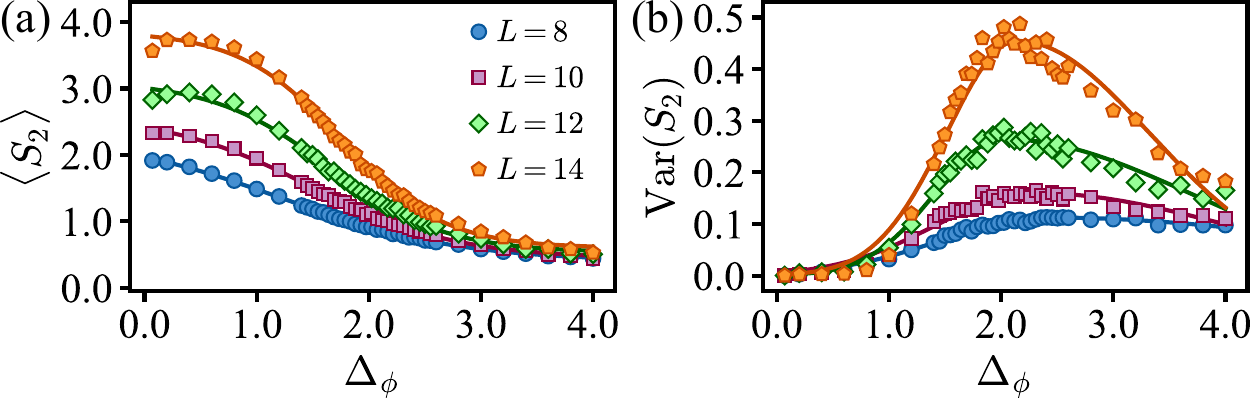}
	\caption{(a) Mean half-system block entanglement $\langle S_{2} \rangle$, averaged over ten eigenstates in the middle of the spectrum, as a function of the disorder strength $\Delta_\phi$ for system sizes up to $L=14$ spins. (b) Variance of the block entanglement over disorder ensembles as a function of $\Delta_\phi$. The maximum grows with $L$ and occurs at $(\Delta_\phi)_{\textrm{max}} \approx 2.17$ for $L=14$. The parameters of the system are the same as in Fig.~\ref{fig:scheme-corre}.}
	\label{fig:entropy}
\end{figure}

We exactly diagonalize the transmon Hamiltonian for system sizes $L \in \{ 8, 10, 12, 14\}$. The analysis is performed on a set of ten eigenstates drawn from the middle of the spectrum, the region expected to be hardest to localize~\cite{Abanin:19}. We will restrict the problem to the largest Hilbert-space sector, corresponding to vanishing total local magnetization~\cite{Bahovadinov:22}. All results are averaged over $10^3$ disorder realizations, from which we obtain the mean R\'enyi entropy $\langle S_{2} \rangle $, shown in Fig.~\ref{fig:entropy}.
Figure~\ref{fig:entropy}(a) shows that, at low disorder, $\langle S_{2} \rangle $ grows with increasing $L$, consistent with an approximate volume-law scaling in the ergodic phase. This trend reverses at large disorder, where $\langle S_{2} \rangle $ saturates to a size-independent value, the hallmark of area-law scaling in the localized phase. Figure~\ref{fig:entropy}(b) shows the variance of the block entanglement, ${\textrm{Var}}(S_{2}) = \langle S_{2}^2 \rangle  - \langle S_{2} \rangle ^2$, as a function of $\Delta_\phi$. Near the critical point, the block entanglement is expected to fluctuate strongly about its mean across disorder realizations, driving a divergence of its variance in the thermodynamic limit~\cite{Kjall:14}. For finite systems, this critical behavior instead appears as a pronounced maximum in ${\textrm{Var}}(S_{2})$, a finite-size precursor of the critical point, which occurs at $(\Delta_\phi)_{\textrm{max}} \approx 2.17$ for $L=14$. This is consistent with the uncorrelated disorder case~\cite{Bahovadinov:22}. 

Despite the success of Rényi entropy as a probe of the MBL transition, resolving stationary, highly excited states of the Hamiltonian remains challenging. To circumvent this, we introduce below an MBL characterization based on quantum dynamics.

\paragraph{Quantum Dynamics.} 

To quantify the extent to which an initial local density profile is preserved under unitary time evolution, we introduce a dynamical distance metric. 
The instantaneous deviation of the time-evolved profile from its initial state is quantified by the profile distance,
\begin{equation}
	D(t) = \sum_{i=1}^{L} \left| n_i(t) - n_i(0) \right|,
	\label{eq:profile_distance}
\end{equation}
with $n_i(t) = \langle \psi(t) | \hat{n}_i | \psi(t) \rangle$. Note that $0\le D(t) \le L$, with $D(t) = 0$ if the initial density configuration is perfectly preserved, whereas $D(t) > 0$ when the initial local structure is altered by the dynamics. For a system with $N_{\mathrm{exc}}$ initially excited qubits, we define the filling factor $\nu=N_{\mathrm{exc}}/L$. In the ergodic phase, thermalization is expected to asymptotically occur for uniform density $n_i(t\rightarrow \infty)=\nu$, $\forall$ $i$. In this case,  we obtain $D(t) \to D_{\mathrm{erg}} = 2\nu (1 - \nu)L$.  

Therefore, we define a normalized disorder-averaged local-memory parameter as
\begin{equation}
	\langle M(t) \rangle := 1 - \frac{\langle D(t) \rangle}{D_{\mathrm{erg}}} .
	\label{eq:local_memory_definition}
\end{equation}
For half-filling ($\nu = 1/2$), the allowed range is $-1 \le \langle M(t) \rangle \le 1$, although in practice the observed dynamics typically span a narrower interval. The upper bound $\langle M(t) \rangle \to 1$ corresponds to a system that retains (or returns to) its initial local profile, a perfect-memory state, while $\langle M(t) \rangle \to 0$ characterizes the ergodic phase, where the local density relaxes to the uniform value $\nu$. Values $\langle M(t) \rangle < 0$ can occur transiently and signal an anti-correlated configuration, in which the local profile has evolved away from the initial state more than it would under simple thermalization, with the extreme case being a complete population swap between initially occupied and empty sites.

From an experimental standpoint, unlike non-local metrics such as many-body entanglement entropy, evaluating $\langle M(t) \rangle$ requires only single-qubit local population measurements, $\langle \hat{n}_i \rangle$, on each qubit in the system. This makes it a highly advantageous observable for noisy intermediate-scale quantum (NISQ) processors, such as superconducting transmon platforms, since it drastically reduces the measurement overhead and is directly accessible via both standard projective and dispersive single-qubit readouts.

\begin{figure}[t!]
	\centering
	\includegraphics[width=\columnwidth]{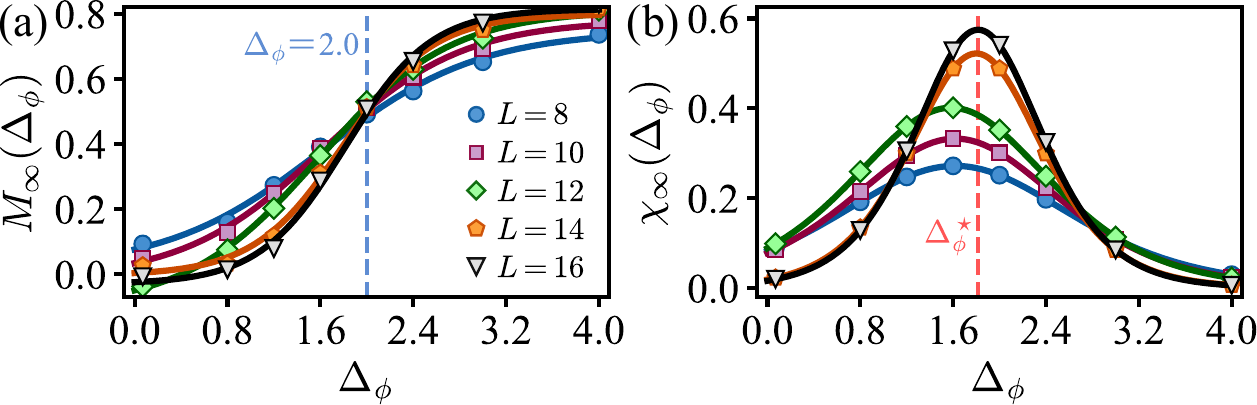}
	\caption{(a) Disorder-averaged late-time local-memory parameter
		$\langle M\rangle$ as a function of the disorder strength
		$\Delta_\phi$ for $L\in\{8,10,12,14,16\}$. The memory is evaluated for a time $\tau_L\simeq 864$, corresponding to
		$t \simeq 320~\mathrm{ns}$, with
		$\tau_L=g_{\mathrm{nn}}t$. The curves exhibit a scale-invariant crossing point at $\Delta_\phi \simeq 2$. (b) Numerical derivative $d\langle M \rangle /d\Delta_\phi$ versus $\Delta_\phi$. The peak of the derivative scales with system size $L$, reaching $\left(d\langle M \rangle /d\Delta_\phi\right)_{\max} \simeq 0.58$ for $L=16$.}
	\label{fig:dynamics}
\end{figure}

To evaluate this memory parameter, we prepare the system in the initial state $\ket{\psi_0} = \ket{e}^{\otimes L/2}\ket{g}^{\otimes L/2}$, where the $L/2$ first qubits are fully-excited. In this case, we get $D_{\mathrm{erg}} = L/2$. If the system undergoes ergodic dynamics, the asymptotic information spreading leads to $n_i(t) \to 1/2$ and $\langle M(t) \rangle \to 0$. Conversely, if the system is in a localized phase, the evolved profile remains close to the initial state, and $D(t) \ll D_{\mathrm{erg}}$.

We measure time in units of the mean nearest-neighbor coupling
of the corresponding clean array,
$g_{\mathrm{nn}}
=(L-1)^{-1}\sum_{i=1}^{L-1}
g_{i,i+1}|_{\boldsymbol{\phi}=0}$ by defining $\tau_L=g_{\mathrm{nn}}t$.
Here, $g_{\mathrm{nn}}$ is expressed as an angular frequency,
consistently with $\hbar=1$, and
$g_{\mathrm{nn}}/(2\pi) \approx 0.43~\mathrm{GHz}$
for the system sizes considered.

The long-time disorder-averaged local memory parameter $\langle M \rangle$ is shown as a function of the disorder strength $\Delta_\phi$ in Fig.~\ref{fig:dynamics}. 
The numerical simulation considers domain walls for system sizes spanning $L = 8$ to $L = 16$ transmons, averaged over $1,000$ disorder realizations for $L \le 14$ and $500$ realizations for $L = 16$. 
All curves exhibit an inflection point with a shared crossing point near $\Delta_\phi \simeq 2$ [Fig.~\ref{fig:dynamics}(a)]. 
By examining the numerical derivative $\chi_{\infty} (\Delta_\phi) := d\langle M_{\infty} \rangle /d\Delta_\phi$ [Fig.~\ref{fig:dynamics}(b)], we characterize the sharpening of this inflection point with increasing $L$. 
For the largest system size, $L=16$, the maximum dynamical
response occurs at $\Delta_\phi^\star\simeq1.82$, with $\chi_{\infty}^{\max}\simeq0.57$. Note that $\Delta_\phi^\star$ is lower than the finite-size precursor $(\Delta_\phi)_\textrm{max}$ found by the R\'enyi entropy for the middle of the spectrum. Since these are the hardest  states to localize, the disorder-averaged local-memory parameter provides an estimate for the MBL critical point for the state $ \ket{e}^{\otimes L/2}\ket{g}^{\otimes L/2}$ that is fully consistent with entanglement entropy diagnostics.

\paragraph{Conclusion.} We have characterized the MBL transition in a superconducting transmon system subject to correlated disorder, a feature intrinsic to the transmon's physical design. We showed that correlated disorder leaves the MBL transition in close correspondence with the uncorrelated case, indicating that the transmon platform is robust against correlations in its random couplings and fields. This robustness is established through the half-system block entanglement variance. We also introduced a quantum-dynamical approach based on a long-time disorder-averaged memory parameter, showing that the same critical behavior is captured, providing an analysis that is readily accessible in experimental realizations.

\begin{acknowledgments}
	\paragraph{Acknowledgments.} 
	T.R.G.S. acknowledges financial support by Coordena\c{c}\~ao de Aperfei\c{c}oamento de Pessoal de N\'{\i}vel Superior (CAPES). 
	A.C.S. is supported by the Comunidad de Madrid through the program Ayudas de Atracción de Talento Investigador ``César Nombela", under Grant No. 2024-T1/COM-31530 (Project SWiQL).
	M.S.S. is supported by Conselho Nacional de Desenvolvimento Cient\'{\i}fico e Tecnol\'ogico (CNPq) (grant number 303836/2024-5).
\end{acknowledgments}

\textbf{Data Availability.}
The processed datasets, Python scripts, and notebooks used to reproduce the theoretical figures in the main text and Supplemental Material are available in the GitHub repository~\cite{github_repo}. Full production runs require high-performance computing resources and are not included in the repository.


%



\onecolumngrid
\newpage

\begin{center}
	{\large{ {\bf Supplemental Material for: \\ Many-Body Localization Induced by Correlated Disorder in Interacting Superconducting Qubits}}}

\vskip\baselineskip{Thiago R. Gir\~ao Souza,$^{1,{\color{blue}\ast}}$ Andreia Saguia,$^{1,{\color{blue}\dagger}}$ Alan C. Santos,$^{2,3,{\color{blue}\ddagger}}$ and Marcelo S. Sarandy$^{1,{\color{blue}\S}}$}

\vskip0.5\baselineskip{\small{\it$^{1}$\UFFSM}\\
	{\it $^{2}$\QuARC}
	\\
	{\it $^{3}$\CSIC}
}

\vskip\baselineskip{$^{\color{blue}\ast}$thiagorgs@id.uff.br, ~~~ $^{\color{blue}\dagger}$asaguia@id.uff.br, ~~~ $^{\color{blue}\ddagger}$alan.santos@csic.es, ~~~ $^{\color{blue}\S}$msarandy@id.uff.br}
\end{center}

\vspace{1cm}

\appendix

\setcounter{equation}{0}
\setcounter{figure}{0}
\setcounter{table}{0}

\renewcommand{\theequation}{S\arabic{equation}}
\renewcommand{\thefigure}{S\arabic{figure}}
\renewcommand{\bibnumfmt}[1]{[S#1]}
\renewcommand{\citenumfont}[1]{S#1}

\twocolumngrid

\section{ENGINEERING DISORDER}

As discussed in the main text, the Hamiltonian of the transmon device is
\begin{align}
	\hat{H}_{\mathrm{nloc}}
	=
	\sum_{m=1}^{N}
	\hbar\omega_{m}(\boldsymbol{\phi})
	\hat{\sigma}^{+}_{m}\hat{\sigma}^{-}_{m}
	+
	\sum_{m,n<m}^{N}
	\hbar g_{nm}(\boldsymbol{\phi})
	\left(
	\hat{\sigma}^{+}_{n}\hat{\sigma}^{-}_{m}
	+
	\hat{\sigma}^{+}_{m}\hat{\sigma}^{-}_{n}
	\right),
\end{align}
where $\boldsymbol{\phi}=\{\phi_1,\ldots,\phi_N\}$ denotes the set of local external fluxes applied to the qubits. Since both the qubit frequencies $\omega_m(\boldsymbol{\phi})$ and the couplings $g_{nm}(\boldsymbol{\phi})$ depend on the applied fluxes, the introduction of  frequency disorder through random values of $\phi_n$ inevitably induces fluctuations in the couplings as well.

The Hamiltonian effectively used to describe the theoretical disordered model is
\begin{align}
	\hat{H}_{\mathrm{dis}}
	=
	\sum_{m=1}^{N}
	\hbar h_m
	\hat{\sigma}^{+}_{m}\hat{\sigma}^{-}_{m}
	+
	\sum_{m,n<m}^{N}
	\hbar J_{nm}
	\left(
	\hat{\sigma}^{+}_{n}\hat{\sigma}^{-}_{m}
	+
	\hat{\sigma}^{+}_{m}\hat{\sigma}^{-}_{n}
	\right),
\end{align}
where $h_m\in[-\Delta J,\Delta J]$, and $J$ defines the characteristic interaction energy scale. In this appendix we describe how the quantities $\Delta$ and $J$ are obtained from the experimentally accessible Hamiltonian $\hat{H}_{\mathrm{nloc}}$.

To exemplify our strategy, we consider the qubit frequency and coupling strength are given by
\begin{align}
	\hbar \omega_m
	&=
	\sqrt{8E_{C,m}E_{J,m}( \phi_m)}
	-E_{C,m},
	\\
	\hbar g_{nm}
	&=
	\sqrt{2}\,E_{C,nm}
	\left(
	\frac{
		E_{J,n}( \phi_n)
		E_{J,m}( \phi_m)
	}{
		E_{C,n}E_{C,m}
	}
	\right)^{1/4}.
\end{align} 
where the Josephson energy is defined from~\cite{Koch:07}, and given by $E_{J}(\phi) = E_{J,0} [\cos^{2} (\pi \phi) + \alpha^{2}\sin^{2} (\pi \phi)]^{1/2}$, where $\alpha \leq 1.0$ is the junction asymmetry parameter for tunable transmons. So, we begin by defining the zero-flux qubit frequencies,
\begin{align}
	\omega_{0,m}
	=
	\omega_m(\boldsymbol{\phi}=0),
\end{align}
which correspond to the maximum frequencies attainable by each qubit. Due to fabrication imperfections, these frequencies are generally not identical. We therefore define the largest zero-flux frequency as
\begin{align}
	\omega_{\mathrm{max}}
	=
	\max_m\{\omega_{0,m}\}.
\end{align}

Next, let $\phi_0$ denote the maximum value that any randomly generated flux can assume, such that $\phi_n\in[0,\phi_0]$, with $\phi_0\leq 0.5$. The minimum frequency of each qubit within this interval is then $\omega_{\mathrm{min},m}
=
\omega_m(\boldsymbol{\phi}=\phi_0 )$. Moreover, we define the absolute minimum frequency of the device as
\begin{align}
	\omega_{\mathrm{min}}
	=
	\min_m\{\omega_{\mathrm{min},m}\}.
\end{align}
Now, because the maximum frequency range accessible for the chosen value of $\phi_0$ is therefore
$\delta\omega(\phi_0)=\omega_{\mathrm{max}}-\omega_{\mathrm{min}}$, while the corresponding central frequency is
\begin{align}
	\bar{\omega}(\phi_0)
	=
	\frac{\omega_{\mathrm{max}}+\omega_{\mathrm{min}}}{2}.
\end{align}	
We can define the disorder amplitude as one half of the accessible frequency range,
\begin{align}
	\delta(\phi_0)
	=
	\frac{\omega_{\mathrm{max}}-\omega_{\mathrm{min}}}{2}.
\end{align}
Consequently, the on-site energies satisfy
\begin{align}
	h_m\in[-\delta(\phi_0),\delta(\phi_0)],
\end{align}
showing explicitly that the disorder amplitude is controlled by the maximum applied flux $\phi_0$.

\begin{figure*}[t!]
	\centering
	\includegraphics[width=\linewidth]{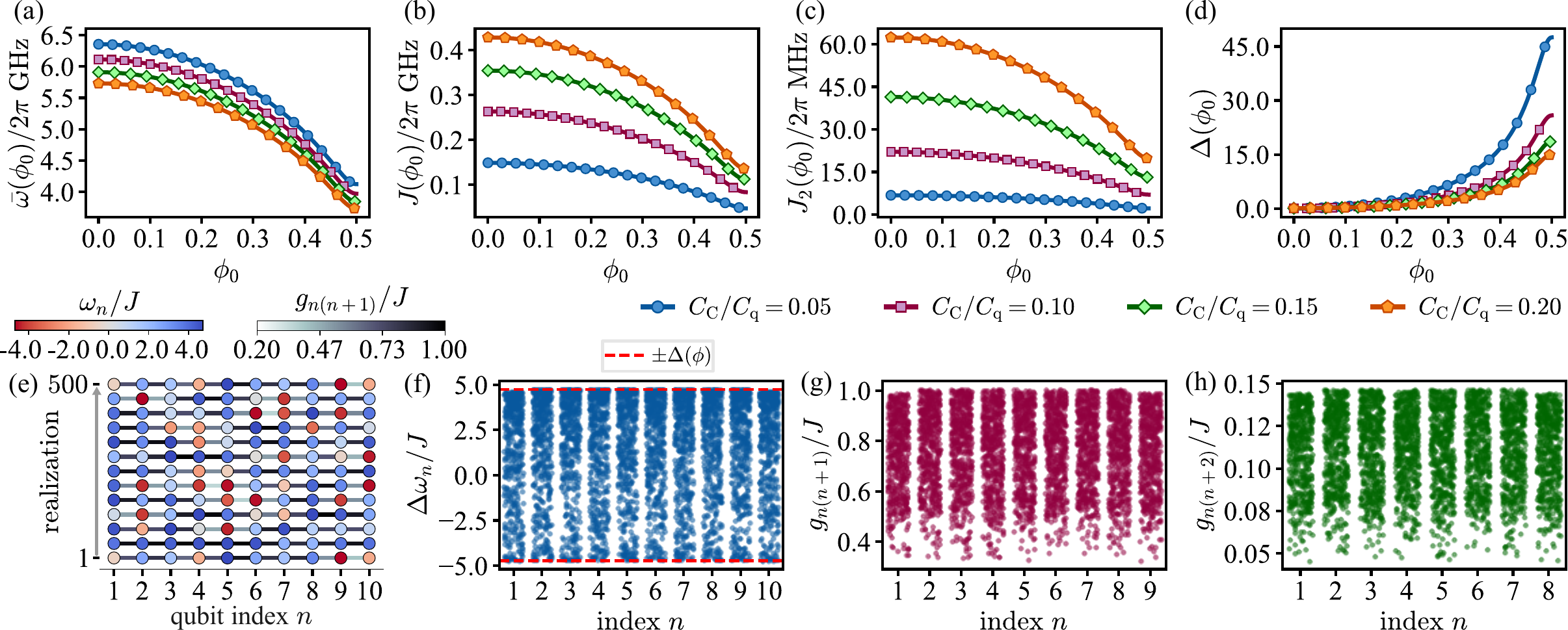}
	\caption{(a) Mean value of the frequency, (b) nearest-neighbor, (c) next-nearest-neighbor coupling strength and (d) \textit{in situ} amplitude of disorder as function of the external flux $\phi_{0}$ for different values of the ratio $C_\mathrm{C}/C_\mathrm{q}$.
		(e) Diagram showing the \textit{in situ} disorder in frequency for the qubits (circles) and couplings (edges) for different realizations, with the complete values of all realizations for (f) qubit frequencies, (g) nearest-neighbor, (h) next-nearest-neighbor interactions.
		We consider a system of $10$ qubits, with experimentally feasible parameters $C_{\mathrm{S},n} = 100$~fF, $C_{\mathrm{C}} = 0.2C_{\mathrm{S},n}$ and $E_{\mathrm{J},n} = 2 \times 10^{-23}$~J~\cite{Ding:23,Wu:24}. The plots from (e) to (h) considered $500$ realizations and we set each $\phi_n$ randomly in the interval $[0,0.5]$.}
	\label{fig:SM-parameters}
\end{figure*}

To obtain a dimensionless disorder strength, we define the average nearest-neighbor coupling as
\begin{align}
	J(\phi_0)
	=
	\frac{1}{N-1}
	\sum_{n=1}^{N-1}
	g_{n,n+1}(\phi_0),
\end{align}
where the couplings are evaluated for the same value of $\phi_0$. Then, the dimensionless disorder parameter is 
\begin{align}
	\Delta(\phi_0)
	=
	\frac{\delta(\phi_0)}{J(\phi_0)}.
\end{align}
This construction provides a direct mapping between the experimentally accessible control parameter $\phi_0$ and the dimensionless disorder strength $\Delta$. Although varying the local fluxes also induces fluctuations in the couplings $g_{nm}$, these variations remain substantially smaller than the corresponding changes in the qubit frequencies over the operating range considered here. Consequently, the dominant effect of the applied flux is the realization of controllable on-site and coupling disorder simultaneously.

The dependence of the system parameters on the applied flux is summarized in Fig.~\ref{fig:SM-parameters}. 
As shown in Fig.~\SubFig{fig:SM-parameters}{a}, the mean qubit frequency $\bar{\omega}(\phi_0)$ exhibits a pronounced nonlinear dependence on the external flux, with a minimum occurring around $\phi_0\simeq 0.5$. 
This behavior originates from the flux dependence of the Josephson energy, which directly modifies the effective onsite energies of the qubits. 
Importantly, the accessible frequency range spans several times the characteristic coupling scale $J(\phi_0)$, allowing the realization of a broad range of effective onsite disorder strengths. 
The different curves correspond to increasing values of the capacitive participation ratio $C_C/C_q$, which primarily affects the interaction energy scale while preserving the overall flux dependence.

The corresponding variation of the nearest-neighbor coupling is presented in Fig.~\SubFig{fig:SM-parameters}{b}. 
Unlike the qubit frequencies, the dependence of $J(\phi_0)$ on $\phi_0$ is considerably weaker, leading only to a moderate reduction of the coupling strength near the half-flux point. 
For all values of $C_C/C_q$ considered, the relative variation of $J(\phi_0)$ remains substantially smaller than the variation of $\omega_m$, confirming that the applied flux predominantly generates onsite disorder rather than strong coupling disorder. 

We also consider the analysis for the next-to-nearest-neighbor interactions, defined as
\begin{align}
	J_{2}(\phi_0)
	=
	\frac{1}{N-2}
	\sum_{n=1}^{N-2}
	g_{n,n+2}(\phi_0).
\end{align} 
The coupling $J_{2}(\phi_0)$ is shown in Fig.~\SubFig{fig:SM-parameters}{c}. Notice that it follows the same qualitative trend as the nearest-neighbor couplings, with a weaker overall magnitude. 
The ratio between the nearest- and next-to-nearest-neighbor interactions remains approximately constant over the full flux range, indicating that the flux control preserves the spatial structure of the interaction network. 
Therefore, the external flux acts mainly as a tunable disorder knob for the onsite energies while maintaining the underlying connectivity of the quantum circuit.

As displayed in Fig.~\SubFig{fig:SM-parameters}{d}, $\Delta(\phi_0)$ is strongly enhanced around $\phi_0\simeq0.5$, where the qubit frequencies become more sensitive to small flux variations. 
The peak value of $\Delta$ increases with decreasing $C_C/C_q$, reflecting the smaller coupling scale used for normalization, and reaches values well beyond unity. 
This regime corresponds to strongly disordered configurations in which the onsite energy fluctuations dominate over the hopping processes, enabling the exploration of localization phenomena.

To further demonstrate the controllability and reproducibility of the generated disorder, we show representative realizations obtained from random local flux configurations in Figs.~\SubFig{fig:SM-parameters}{e--h}. 
The resulting frequency shifts $\Delta\omega_n/J$ cover a broad distribution within the targeted disorder window, while the dashed boundaries indicate the imposed disorder range. 
The corresponding nearest-neighbor and next-nearest-neighbor couplings remain narrowly distributed around their average values, with fluctuations significantly smaller than those of the onsite energies. 
These results confirm that the flux-controlled architecture provides an efficient route to engineering tunable disorder landscapes, where both diagonal and weak correlated off-diagonal disorder are naturally generated by the same experimental control mechanism.

\section{CORRELATED DISORDER ANALYSIS} \label{Sect:Corre}

The data shown in Figs.~\SubFig{fig:SM-parameters}{e--h} suggests that the residual flux dependence of the interactions provides an experimentally relevant source of correlated off-diagonal disorder, which is naturally included in the model without requiring additional control parameters. So, to quantify this correlation, without loss of generality, let us consider small fluctuations of the external flux around their mean values, namely $ \phi_m = \bar{\phi}_m+\delta\phi_m$ and $ \phi_n
=\bar{\phi}_n+\delta\phi_n$, where $\delta\phi_m$ and $\delta\phi_n$ are random fluctuations describing the disorder.

Since both quantities depend on the same fluctuating variable
\( \phi_m\), the disorder induces correlations between
\(\omega_m\) and \(g_{nm}\). In this case, we expand both quantities to first order in the flux fluctuations,
\begin{align}
	\delta\omega_m
	&=
	\frac{\partial \omega_m}{\partial \phi_m}
	\delta\phi_m,
	\quad 
	\delta g_{nm}
	=
	\frac{\partial g_{nm}}{\partial  \phi_m}
	\delta\phi_m
	+
	\frac{\partial g_{nm}}{\partial  \phi_n}
	\delta\phi_n.
\end{align}

From the above equation, we compute the covariance between the frequency and coupling fluctuations defined as $\mathrm{Cov}(\omega_m,g_{nm})=\left\langle\delta\omega_m\,\delta g_{nm}\right\rangle$, with denoting the $\langle \bullet \rangle$ statistical average over many realizations of disorder. Substituting the linearized expressions yields
\begin{align}
	\mathrm{Cov}(\omega_m,g_{nm})
	&=
	\frac{\partial \omega_m}{\partial  \phi_m}
	\frac{\partial g_{nm}}{\partial  \phi_m}
	\left\langle
	(\delta\phi_m)^2
	\right\rangle
	+
	\frac{\partial \omega_m}{\partial  \phi_m}
	\frac{\partial g_{nm}}{\partial  \phi_n}
	\left\langle
	\delta\phi_m\delta\phi_n
	\right\rangle.
\end{align}

Because the local fluxes are implemented through different flux lines it is reasonable to impose they are statistically independent---i.e. $\left\langle \delta\phi_m\delta\phi_n \right\rangle = 0$---which allows us to conclude that
\begin{align}
	\mathrm{Cov}(\omega_m,g_{nm})
	&=
	\frac{\partial \omega_m}{\partial  \phi_m}
	\frac{\partial g_{nm}}{\partial  \phi_m}
	\left\langle
	(\delta\phi_m)^2
	\right\rangle
\end{align}

\section{MBL TRANSITION VIA VON NEUMANN ENTROPY}

\begin{figure}[t]
	\centering
	\includegraphics[width=\columnwidth]{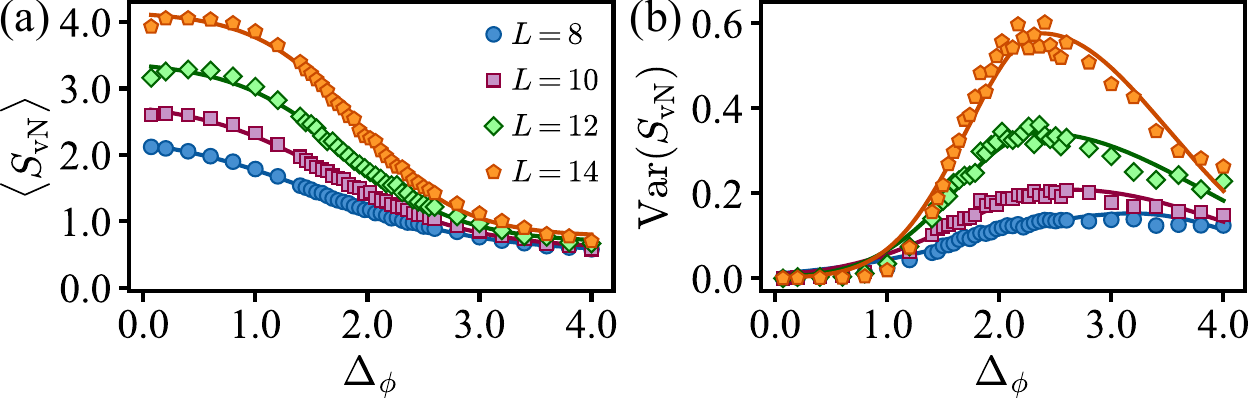}
	\caption{(a) Mean half-system block entanglement $\langle S_{\mathrm{vN}} \rangle$, averaged over ten eigenstates in the middle of the spectrum, as a function of the disorder strength $\Delta_\phi$ for system sizes up to $L=14$ spins. (b) Variance of the block entanglement over disorder ensembles as a function of $\Delta_\phi$. The maximum grows with $L$ and occurs at $(\Delta_\phi)_{\textrm{max}} \approx 2.41$ for $L=14$. The parameters of the system are as described in Fig. 1 of the main text.}
	\label{fig:entropyVn}
\end{figure}

As discussed in the main text, the results obtained with 2nd-order R\'enyi entropy can be confirmed through the von Neumann entropy analysis. To demonstrate this, given an energy eigenstate $|\psi\rangle$ and a bipartition of the transmon array into two contiguous halves $A$ and $B$, the entanglement between $A$ and $B$ is quantified by the von Neumann entropy $S_{\mathrm{vN}}$ of either block,
\begin{equation}
	S_\mathrm{vN} = -\text{Tr}\left(\rho_A \log_2 \rho_A\right) = -\text{Tr}\left(\rho_B \log_2 \rho_B\right),
	\label{vonNeumann}
\end{equation}
where $\rho = |\psi\rangle\langle\psi|$ and $\rho_A = \text{Tr}_B\,\rho$, $\rho_B = \text{Tr}_A\,\rho$ are the reduced density matrices of blocks $A$ and $B$, respectively.

Figure~\SubFig{fig:entropyVn}{a} shows that, at low disorder, $\langle S_{\mathrm{vN}} \rangle $ grows with increasing $L$, consistent with an approximate volume-law scaling in the ergodic phase. This trend reverses at large disorder, where $\langle S_{\mathrm{vN}} \rangle $ saturates to a size-independent value, the hallmark of area-law scaling in the localized phase. Figure~\SubFig{fig:entropyVn}{b} shows the variance of the block entanglement, ${\textrm{Var}}(S_{\mathrm{vN}}) = \langle S_{\mathrm{vN}}^2 \rangle  - \langle S_{\mathrm{vN}} \rangle ^2$, as a function of $\Delta_\phi$. Near the critical point, the block entanglement is expected to fluctuate strongly about its mean across disorder realizations, driving a divergence of its variance in the thermodynamic limit~\cite{Kjall:14}. For finite systems, this critical behavior instead appears as a pronounced maximum in ${\textrm{Var}}(S_{\mathrm{vN}})$, a finite-size precursor of the critical point, which occurs at $(\Delta_\phi)_{\textrm{max}} \approx 2.41$ for $L=14$. This is consistent with the uncorrelated disorder case~\cite{Bahovadinov:22}. 

\newpage

\end{document}